\documentclass[proof]{WileyASNA-v1}

\articletype{Article Type}%

\received{xx April 2022}
\revised{xx June 2022}
\accepted{xx June 2022}

\begin{document}

\title{Abundance of radioactive technetium in Przybylski's star revisited}

\author[1,2,3]{Sergei M. Andrievsky*}
\author[4]{Sergey A. Korotin}
\author[2]{Klaus Werner}

\authormark{Sergei M. Andrievsky et al.} 

\address[1]{\orgdiv{Astronomical Observatory}, \orgname{Odessa National University of the Ministry of 
Education and Science of Ukraine}, \orgaddress{\state{Shevchenko Park, 65014, Odessa}, \country{Ukraine}}}

\address[2]{\orgdiv{Institut f\"{u}r Astronomie und Astrophysik}, \orgname{Kepler Center for 
Astro and Particle Physics, Universit\"{a}t T\"{u}bingen}, \orgaddress{\state{Sand 1, 72076 T\"{u}bingen}, 
\country{Germany}}}

\address[3]{\orgdiv{GEPI}, \orgname{Observatoire de Paris, Universit\'e PSL, CNRS}, 
\orgaddress{\state{ 5 Place Jules Janssen, F-92190 Meudon}, \country{France}}}

\address[4] {\orgdiv{Physics of stars department}, \orgname{Crimean Astrophysical Observatory}, 
\orgaddress{\state{Nauchny 298409}, \country{Crimea}}}


\corres{*S.~M.~Andrievsky, Astronomical Observatory, Odessa National University of the Ministry of 
Education and Science of Ukraine, Shevchenko Park, 65014, Odessa, Ukraine \email{andrievskii@ukr.net}}

\abstract{We have searched for lines of the radioactive element technetium (Tc) in the spectrum of
Przybylski's star (HD~101065). The nuclei of this chemical element are formed in the slow process
of capturing thermalized neutrons. The possible lines of Tc~I are heavily blended. We have synthesized 
the profile of one resonance line at 4297.06 \AA~, which is also a part of the complex blend, and we arrived 
at a decision that it is not visible in the spectrum (as was first noted by Ryabchikova), casting doubt 
on the existence of technetium in the atmosphere of the Przybylski's star. Therefore, based on our calculated 
combined profile, which has been adjusted to the observed blend profile at $\approx$ 4297.2 \AA~ 
(that may possibly contain the resonance technetium line 4297.06 \AA), we reduce the maximum technetium 
abundance to $\log\epsilon$(Tc/H) = 2.5. This value can be considered only as an upper limit of 
the technetium abundance in the Przybylski's star.}


\keywords{Peculiar stars: atmospheres: Przybylski's star}

\maketitle
   
\section{Introduction}

Technetium lines were first detected by \cite{Merrill1952} in the spectra of several type S
ZrO stars. Further studies of S-stars showed that some of them display strong absorption 
features in their spectra near the expected positions of technetium lines, while others do not. 
It is believed that stars with technetium are at the stage of AGB stars with thermal pulses. 
$s$-process elements, including technetium, may appear on the surface of the star as a result
of the third dredge-up episode. Most recent papers devoted to the study of 23 S-stars have been 
published by \cite{Shetyeetal2020} and \cite{Shetyeetal2021}. The authors point out that the 
4262.27 \AA~ Tc line is the best line reproduced by spectrum synthesis. Fig. 1 from the latter 
paper shows how complex the spectra of such cool stars of very low gravity are. Indeed, 
a synthetic spectrum calculation for a star with values of 
temperature, gravity, metallicity and $s$-process element abundances typical for S-stars shows 
a strong blend comprising  the possible Tc~I 4262.27 \AA~ line, which consists of many components. 
Some of these components (V~I and Nd~II) have wavelengths very close to the Tc~I line, and close 
residual fluxes. The very strong Gd~II, Gd~I, Nb~I and Cr~I blending lines are slightly 
blueshifted and redshifted compared to Tc~I line, respectively. They also significantly
affect the shape of the common blend. Even this one example shows how insecure the 
technetium abundance obtained from the spectra of very cool stars with low gravity can be. 

The first attempts to identify the lines of technetium isotopes in the spectrum
of the unique Przybylski's star were made by \cite{Cowleyetal2004} and \cite{Bidelman2005}.
\cite{Cowleyetal2004} (see also the web-page of 
Dr. C. Cowley http://www-personal.umich.edu/$\sim$cowley/prz2r.html) reported the presence of
Tc lines in the spectrum of Przybylski's star. \cite{Bidelman2005} carried out a complete line 
identification in the spectrum of the Przybylski's star and reported the presence of lines of 
short- and long-lived isotopes of various chemical elements, including technetium. Nevertheless, 
the results of this great enthusiastic work unfortunately do not provide reliable grounds to 
believe that some of these identified lines belong, in particular, to Tc~I and Tc~II.

\cite{YushGop2004}, in their overview, also mentioned the presence of the Tc lines in the Przybylski's 
star. \cite{Yushchenkoetal2006} performed a more detailed examination of this problem.
The synthetic spectrum technique was applied to fit the 4124.22 \AA~ line of Tc~I. Although
acceptable fit to the observed spectra was not achieved, the authors 
concluded that absolute abundance of technetium should be somewhere between $\log\epsilon$(Tc/H) = 3 
and 4 (on the scale where $\log\epsilon$(H) = 12.0).

More critical consideration of the technetium problem was made by \cite{Ryabchikova2008}. The author 
noted that the spectrum synthesis well reproduces unblended feature at 4124.22 \AA~ with the adopted technetium 
abundance $\log\epsilon$(Tc/H) = 4.0, but the calculated resonance line at 4297.06 \AA~ with such abundance 
appears to be extremely strong, while it is actually invisible in the observed spectrum.

Absence or presence of the short- or observationally long-lived isotopes of some elements in the 
atmosphere of this enigmatic star may shed a light on the processes that lead to its observational 
chemical peculiarities. Therefore we have undertaken an additional attempt to verify the possibility of 
the technetium presence in the atmosphere of this star. This chemical element (Tc, Z = 43) has 
no stable isotopes. On the Earth, $^{99}$Tc can be formed in small  amounts in uranium, thorium,
and molybdenum ores either due to spontaneous fission of the first two elements, or
by neutron captures by the third element. The half-life of $^{99}$Tc isotope is about 210000 years. 

In the stars, the nuclei of $^{99}$Tc and other long-lived technetium isotopes can be  
produced in the $s$-process (the slow neutron capture by the seed nuclei) and further $\beta^{-}$
decays. For instance, the nuclei of $^{99}$Tc can be formed from nuclei of the stable molybdenum 
isotope $^{98}$Mo as follows:
 
$^{98}$Mo + $n$ $\rightarrow$ $^{99}$Mo

$^{99}$Mo $\rightarrow$ $^{99m}$Tc + $e^{-}$ + $\widetilde{\nu}$,

$^{99m}$Tc $\rightarrow$  $^{99}$Tc + $\gamma$    

In the last reaction, the $^{99m}$Tc isomeric state loses energy due to the emission
of the $\gamma$-quantum and forms $^{99}$Tc.

Possibly, another suitable way of the long-lived Tc isotopes formation can be as
follows. $^{97}$Ru nuclei are formed from the nuclei of the observationally stable isotope  
of $^{96}$Ru by neutron captures. Then, in the following chains, the nuclei of two long-lived 
isotopes $^{97}$Tc and $^{98}$Tc (half-life times are about 4.2 million years) can be formed:

$^{97}$Ru = $^{97m}$Tc + $e^{+}$ + $\nu$,

$^{97m}$Tc = $^{97}$Tc + $\gamma$,

$^{97}$Tc + $n$ =  $^{98}$Tc

Here $^{97m}$Tc denotes the nucleus of the $^{97}$Tc isomeric excited state. 

This work is devoted to a revised search for the lines of the unstable element technetium in the spectrum 
of the unique Przybylski's star. 

\section{Spectrum synthesis}

We used an observed spectrum of the Przybylski's star retrieved from the ESO Archive (HARPS, R = 110000, 
S/N = 140, observation date and time 2005-04-03, 05:07:42).

To synthesize the Przybylski's star spectrum in the vicinity of the selected Tc~I lines, we used 
an atmosphere model calculated with the help of the LL models code by \cite{Shulyaketal2004}. This 
code enables one to take into account the line opacities by using individual abundances of elements 
and the possible stratification of chemical elements. Atmosphere parameters and chemical composition 
of the program star are taken from \cite{Shulyaketal2010}. The synthetic spectrum was calculated in 
the local thermodynamical equilibrium approximation using the modified SynthV code (\citealt{Tsymbaletal2019})  
and the above mentioned stratification. The atomic line parameters are from updated Vienna Atomic Line Database 
(\citealt{Ryabchikovaetal2015}). 


Our preliminary test calculations have shown that the strongest Tc lines in the visible part of the spectrum 
are blended in varying degrees by the lines of other elements, therefore most of the Tc lines are unsuitable for 
determining this element abundance. 

The resonance line 4297.06 \AA, which is expected to be strongest from the lines we considered, should be
present in the spectrum, but it is absent. Its calculated profile with the technetium abundance adopted by
\cite{Yushchenkoetal2006}, $\log\epsilon$(Tc/H) $\approx 3.5$, is in a huge discordance with the observed profile 
of this blend. The problem is shown in Fig. 1 (see also discussion in \citealt{Ryabchikova2008}, who made first
such a conclusion). If we decrease the abundance of technetium to $\log\epsilon$(Tc/H) = 2.5, then the overall 
profile of the blend 4297.06 \AA~ can be described more or less satisfactory (Fig. 1). 

Two subordinate lines 4124.22 \AA~ and 4176.28 \AA, which according to our previous test calculations are located 
in relatively blend-free parts of the spectrum (in the sense that there are no other lines listed in modern 
line databases, such as the Vienna Atomic Line Database, \citealt{Ryabchikovaetal2015}, which can be excited at 
the temperature of the program star). 


At the abundance $\log\epsilon$(Tc/H) = 2.5, the two subordinate Tc lines considered above become undetectable 
(see Figs 2--3). The absorption features at 4124.22 \AA~ and 4176.28 \AA~ are more likely to belong to other 
unidentified species, but not  neutral technetium.

Thus, based on the consideration of the resonance line 4297.06 \AA, we can conclude that we have obtained only an 
upper limit on the technetium abundance in the Przybylski's star.

    
\begin{figure} 
\resizebox{\hsize}{!}{\includegraphics{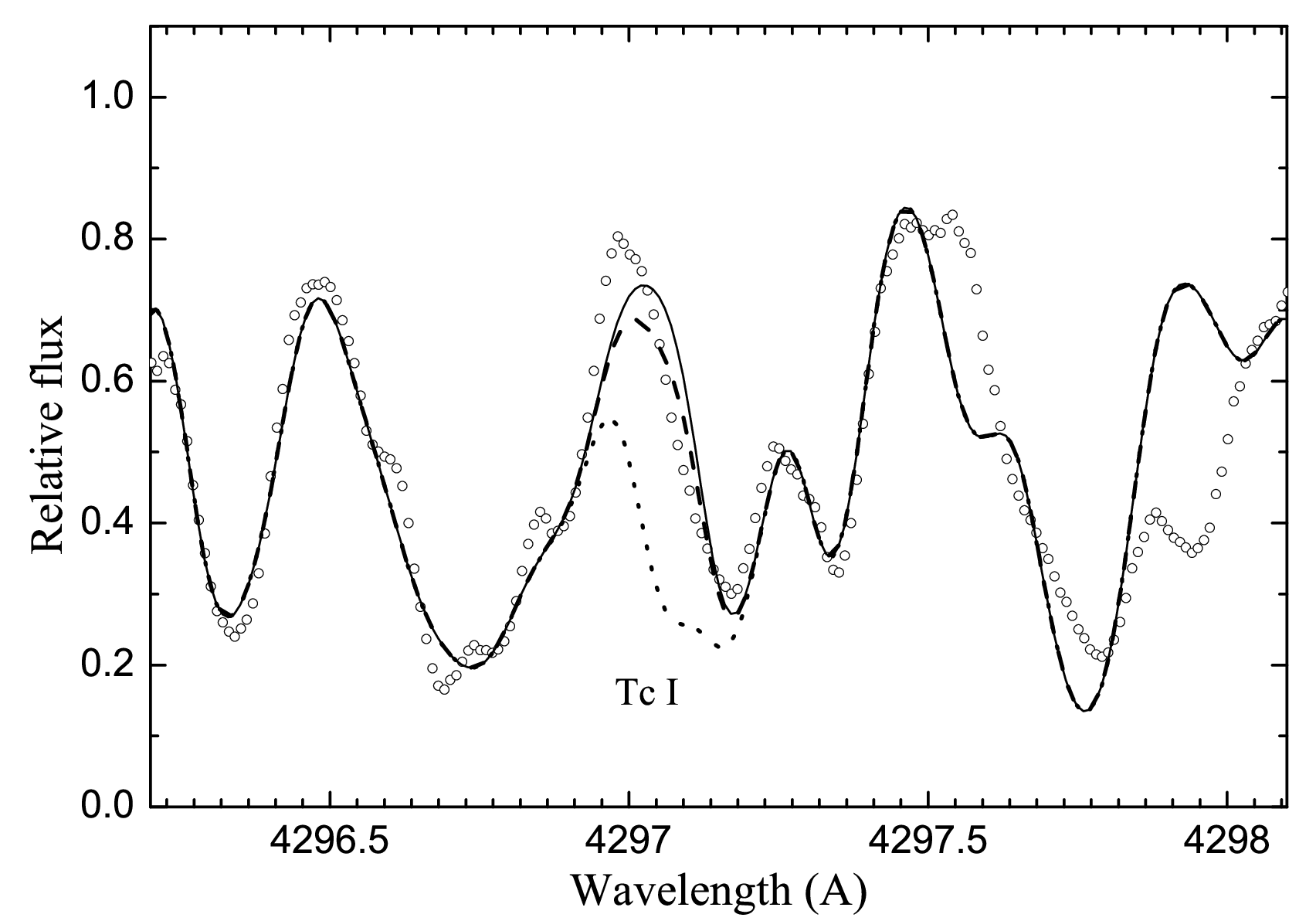}}    
\caption{Fragment of the observed spectrum of the Przybylski's star (open circles), covering
the blend 4297 \AA. Synthetic spectra with technetium abundance  $\log\epsilon$(Tc/H) $\approx 4$ (dotted line), 
and with abundance  $\log\epsilon$(Tc/H) = 2.5 (dashed line) are shown. Continuous line -- no technetium 
in the atmosphere. See text for calculation details}
\label{4297_TcI}
\end{figure}

\begin{figure} 
\resizebox{\hsize}{!}{\includegraphics{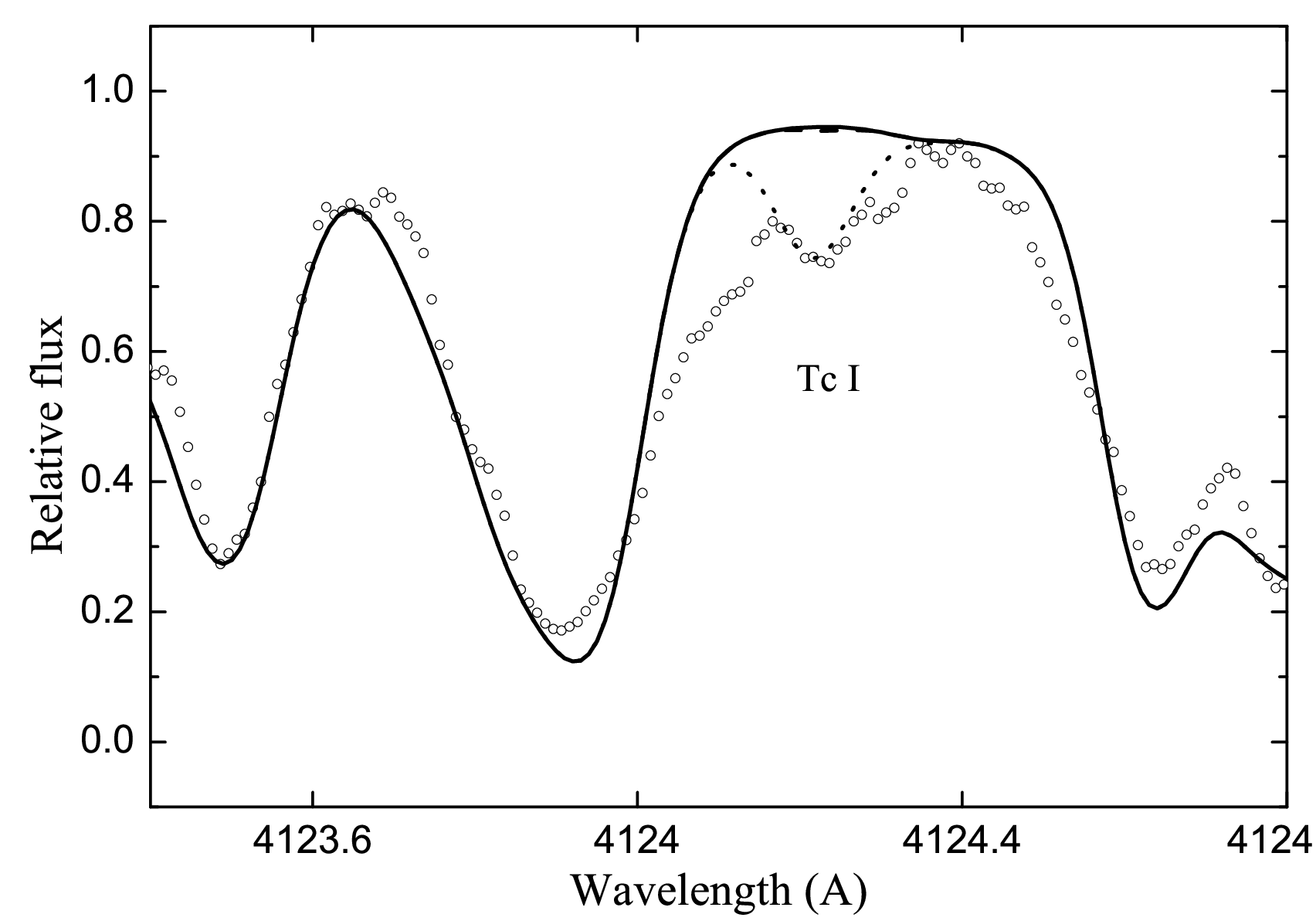}}    
\caption{Same as Fig. 1 but for the feauture at 4124.22 \AA}
\label{4124_TcI}
\end{figure}

\begin{figure} 
\resizebox{\hsize}{!}{\includegraphics{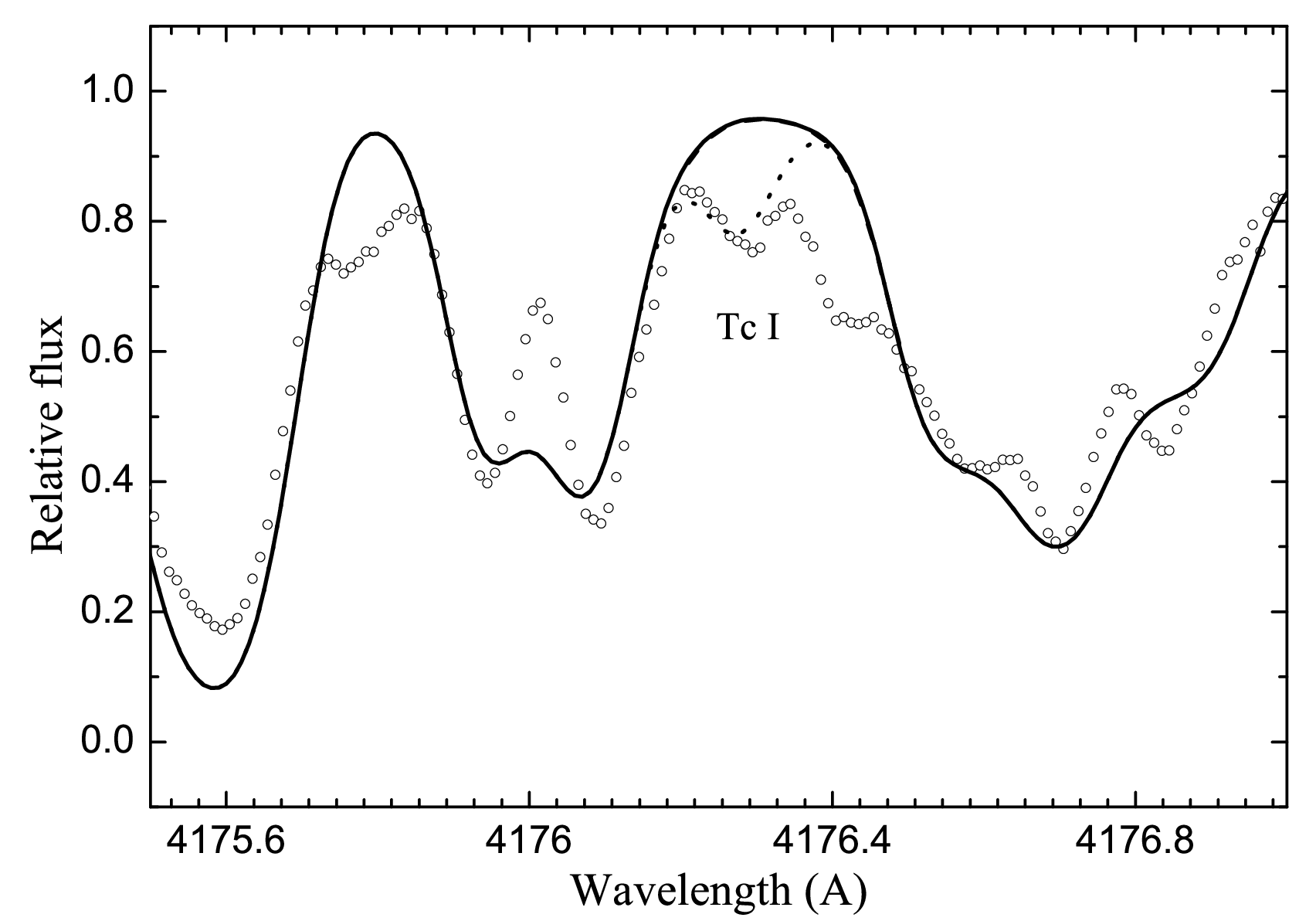}}    
\caption{Same as Fig. 1 but for the 4176.28 \AA~feature}
\label{4176_TcI}
\end{figure}


\section{Discussion and conclusion}

We have made an independent consideration of the possible presence of the rather long-lived
isotopes of technetium in the atmosphere of the unique Przybylski's star. We decreased the possible 
abundance of this element from $\log\epsilon$(Tc/H) $\approx 4$  (see above) to $\log\epsilon$(Tc/H) = 2.5, 
and we consider this value as an upper limit of a possible abundance of this element. Naturally, 
this upper limit means that technetium cannot exist at all in the Przybylski's star atmosphere. 

As was mentioned above, \cite{Cowleyetal2004} and \cite{Bidelman2005} claimed that they detected technetium 
and promethium lines in the spectrum of Przybylski’s star, and they suggested that the presence of these elements 
is perhaps connected to the flare activity of the star. \cite{Goriely2007} elaborated his own physical model, 
which was based on assumption that electromagnetic field accelerates charged particles (protons and $\alpha$-particles, 
in particular) to the energies enough to participate in the nuclear reactions which alter the surface chemical composition.

\cite{Cowleyetal2000} reported a mean magnetic field modulus of 2300 G based on magnetically split rare earth element 
lines with rather large Landé factors, and we verified this with the line of Gd~II at 5749.4 \AA. In the high-resolution 
spectrum we used, the red and blue components are clearly separated by the Zeeman effect, making it possible to measure 
the intrinsic surface field strength with high accuracy. However, 2300 G seems to be too weak for the above mentioned 
high energy processes to be effective. In fact, it was first declared by \cite{BrancazioCameron1967}, and then confirmed 
by \cite{Goriely2007}, who indicated that a magnetic field several orders of magnitude stronger than, for example, 
in Przybylski’s star, is required.


\cite{Andrievsky2022} discusses the hypothesis that considers a neutron star 
as companion of the Przybylski's star, and this neutron star is a $\gamma$-ray source. 
The $\gamma$-ray beam from the neighbouring neutron star irradiates the Przybylski'
star atmosphere causing nuclear reactions that produce free neutrons. 
These neutrons initiate production of the heavy elements. This may explain several 
properties of the Przybylski's star. If the irradiation process is still ongoing, 
then we must detect the presence of the promethium (at least) in the atmosphere of 
this unique star. In \cite{Andrievskyetal2023} we considered the possibility 
of the presence of unstable isotopes of promethium in the atmosphere of Przybylski's star, 
but unfortunately, no unambiguous conclusion about the presence of this element
could be made. If we accept the fact that we do not detect lines of Tc and Pm in the 
Przybylski's star spectrum, then, even not rejecting this hypothesis, we may assume that 
$\gamma$-ray source became inactive, say, 30 million years ago or more (the time, which 
is necessary to decrease the initial technetium abundance by about 100 times or more). 
Such a time is much longer than the lifetime of the Przybylski's star and comparable to
the lifetime of a $\gamma$-ray source.

\subsection*{ACKNOWLEDGEMENTS}

SMA are grateful to the Vector-Stiftung at Stuttgart, Germany, for support within the 
program "2022--Immediate help for Ukrainian refugee scientists" under grants P2022-0063 and P2022-0064,
and DAAD (German Academic Exchange Service). 
SMA would also like to thank the ESO at Garching for their support during his stay there,
which enabled him to perform part of this work.
We are very grateful to our referee for very important comments, which significantly
improved this paper, making it more clear for the reader.

\subsection*{Conflict of interest}

The authors declare no potential conflict of interests.

\end{document}